\documentclass[a4paper,11pt]{article}
\usepackage{pos}

\usepackage{graphicx}
\usepackage{amsmath}
\usepackage{xspace}
\usepackage[numbers,sort&compress]{natbib}
\usepackage{subfig}
\usepackage[shortlabels]{enumitem}

\usepackage{pstricks}
\usepackage{graphicx}
\usepackage{xspace}
\usepackage{hyperref}
\usepackage{lipsum}
\usepackage{multirow}

\newcommand{\hj}{\ensuremath{\rm{H}\,+\,}jet\xspace}
\newcommand{\hjj}{\ensuremath{\rm{H}+2\,}jets\xspace}
\newcommand{\hh}{\ensuremath{\rm{HH}}\xspace}

\newcommand\GeV{\ensuremath{\mathrm{GeV}}\xspace}

\newcommand{\FTapprox}{\ensuremath{\rm{FT}_{\rm approx}}\xspace}
\newcommand{\HTL}{\ensuremath{\rm{HTL}}\xspace}

\newcommand{\gsim}{\;\rlap{\lower 3.5 pt \hbox{$\mathchar \sim$}} \raise 1pt
 \hbox {$>$}\;}
\newcommand{\lsim}{\;\rlap{\lower 3.5 pt \hbox{$\mathchar \sim$}} \raise 1pt
 \hbox {$<$}\;}

\title{
\vskip-3cm{\baselineskip14pt
    \begin{flushright}
     \normalsize \normalfont{TTP24-034, P3H-24-066}
    \end{flushright}} \vskip2.5cm
    
QCD and electroweak corrections for single and double Higgs boson production at the LHC}
 \ShortTitle{QCD and EW corrections for single and double Higgs boson production}

\author*[a]{Hantian Zhang}

\affiliation[a]{Institut f{\"u}r Theoretische Teilchenphysik,
    Karlsruhe Institute of Technology (KIT),\\
 Wolfgang-Gaede Strasse 1, 76128 Karlsruhe, Germany
 }

\emailAdd{hantian.zhang@kit.edu}

\abstract{
In these proceedings, we report recent progress of theoretical predictions for loop-induced Higgs boson production at the Large Hadron Collider~(LHC).
Our contributions include  the next-to-leading order~(NLO) QCD and electroweak corrections for single Higgs boson plus jet production, 
the NLO QCD corrections for single Higgs boson plus two jets production, 
and the NLO electroweak corrections for double Higgs boson production.
}

\FullConference{42nd International Conference on High Energy Physics (ICHEP2024)\\
18-24 July 2024\\
Prague, Czech Republic\\}

\begin{document}
\maketitle

\section{Introduction}
The precise determination of Higgs boson properties
is one of the primary goals in particle physics, particularly in the LHC programme at CERN.
For this purpose, 
single Higgs boson plus jets production and double Higgs boson production are key processes at the LHC,
which can probe the Higgs boson transverse momenta ($p_T^H$) spectrum 
and the trilinear Higgs boson selfcoupling.
They are sensitive to new physics effects beyond the Standard Model~(SM), which are well-studied in the literature such as in Refs.~\cite{LHCHiggsCrossSectionWorkingGroup:2016ypw,Greljo:2017spw,Abouabid:2021yvw,Iguro:2022fel}.
 To distinguish potential new physics effects,
 it is necessary to compute the higher-order QCD and electroweak (EW) radiative corrections
 for single Higgs boson plus jets and double Higgs boson production.
 
 For single Higgs boson plus jet production (\hj),
 the full NLO QCD corrections are computed in Refs.~\cite{Jones:2018hbb,Chen:2021azt,Bonciani:2022jmb}.
In particular, we have performed a dedicated study on top-quark mass ($m_t$) effects in Ref.~\cite{Chen:2021azt}. 
Recent studies on top-quark mass effects on Higgs boson $p_T^H$ spectrum can also be found in the literature such as in Refs.~\cite{Niggetiedt:2024nmp,Czakon:2024ywb}.
 The EW corrections via massless quark loops have been computed in Ref.~\cite{Bonetti:2020hqh}, 
 the corrections induced by a trilinear Higgs coupling in the large-$m_t$ expansion have been calculated in Refs.~\cite{Gao:2023bll}.
 We have computed the first full EW corrections involving top quarks in the large-$m_t$ expansion in Ref.~\cite{Davies:2023npk}. Recently, the beyond-the-SM effects on Higgs boson $p_T^H$ spectrum have been explored in Refs.~\cite{DiNoi:2024ajj,Haisch:2024nzv,Aveleira:2024byi}.

For single Higgs boson plus two jets production (\hjj),
we have computed the first NLO QCD corrections accounting for the top-quark mass effects in Ref.~\cite{Chen:2021azt}.

For double Higgs boson production (\hh), 
the Yukawa-top and Higgs self-coupling corrections are addressed in Refs.~\cite{Borowka:2018pxx,Davies:2022ram,Muhlleitner:2022ijf,Heinrich:2024dnz,Li:2024iio},
the EW corrections involving top quarks are computed analytically in the large-$m_t$ expansion in Ref.~\cite{Davies:2023npk}, 
the factorisable EW corrections are computed analytically in Ref.~\cite{Zhang:2024rix},
and the full EW corrections are computed numerically in Ref.~\cite{Bi:2023bnq}.

In these proceedings, we summarise our contributions for these processes in Refs.~\cite{Chen:2021azt,Davies:2022ram,Davies:2023npk,Zhang:2024rix}.

\section{NLO QCD corrections for \hj and \hjj production}
The computation of NLO QCD corrections for single Higgs boson plus jets production is performed in the \texttt{NNLOJET} framework, which is a parton-level event generators equipped with the antenna subtraction method~\cite{Gehrmann-DeRidder:2005btv}.
The loop-induced matrix elements for \hj and \hjj contributing at the Born level and real corrections are provided by \texttt{OpenLoops2}~\cite{Buccioni:2019sur},
and the matrix elements for virtual corrections of \hj are provided by \texttt{SecDec}~\cite{Borowka:2017idc}.
The input parameters are
$
m_H = 125~\text{GeV}, m_t = 173.055~\text{GeV} ,  v= 246.219~\text{GeV} ,  \sqrt{s} = 13~\text{TeV},
$
where $v$ is the vacuum expectation value and $\sqrt{s}$ is center of mass energy at the LHC. We assume massless light quarks in initial and final states. In our computation, the top-quark mass is renormalised in the on-shell scheme. Renormalisation and factorisation scales are chosen as 
$
\mu_{\rm R,F}=\xi_{\rm R,F} \cdot H_T/2 $ 
with
$ \quad H_T =	\sqrt{m_H^2+p_{\rm{T},H}^2} + \sum\limits_{j} |p_{\rm{T},j}| $,
where the sum includes all final-state partons.
Our central scale corresponds to $\xi_{\rm R,F}=(1,1)$ and we determine scale uncertainties via the standard 7-point factor-2 variations $\xi_{\rm R,F}=(2, 2), (2, 1), (1, 2),$ $ (1, 1), (1,\frac{1}{2}),$ $(\frac{1}{2}, 1),
(\frac{1}{2},\frac{1}{2})$.
We employ the
$p_{\rm{T},j}>30\,\text{GeV}$ cut for \hj production, and 
$p_{\rm{T},j_1}>40\,\GeV \  \mathrm{and} \  p_{\rm{T},j_2}>30\,\GeV$ cuts for \hjj production.

\begin{figure}[t]
\centering
	\includegraphics[width=0.49\textwidth]{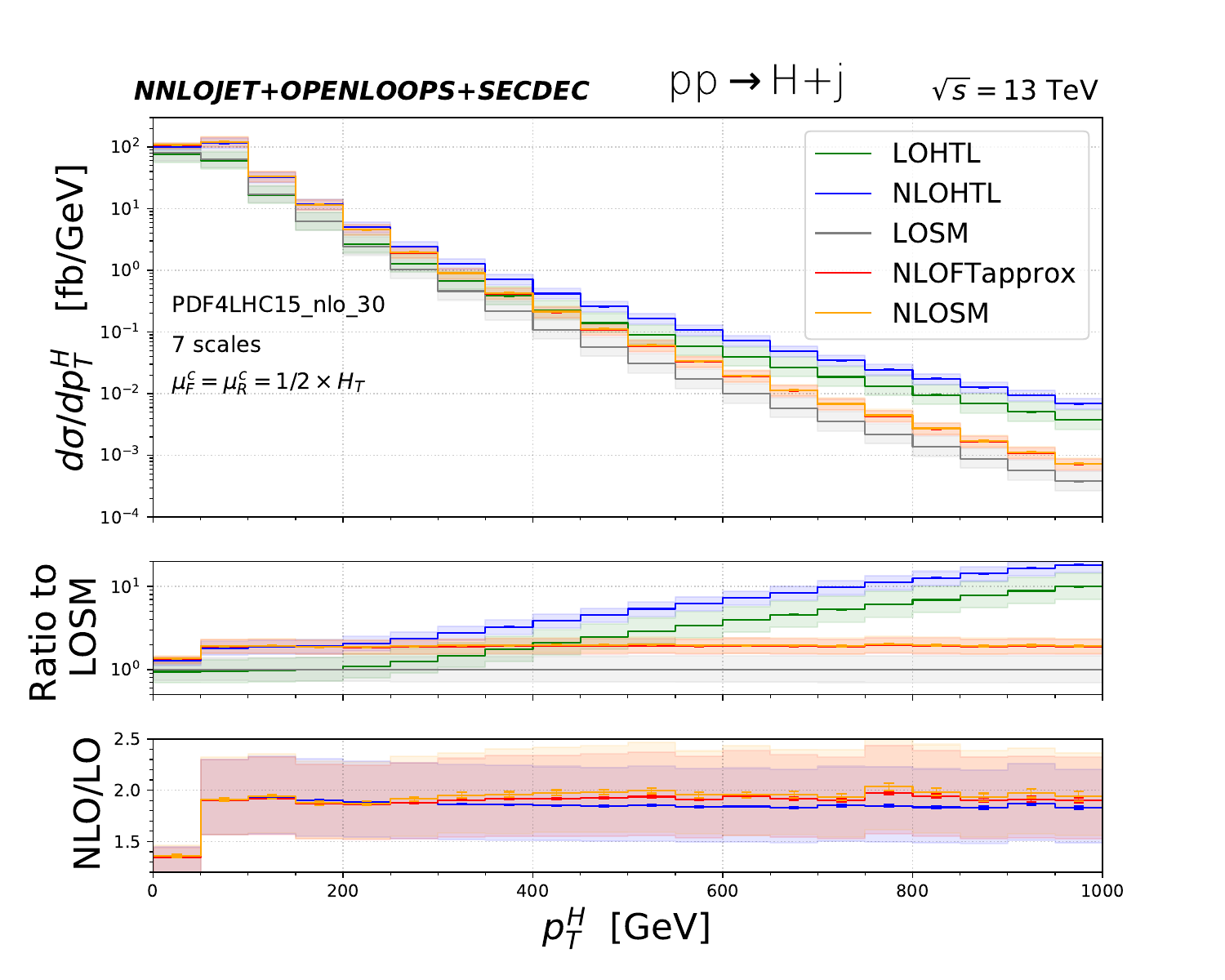}
\includegraphics[width=0.49\textwidth]{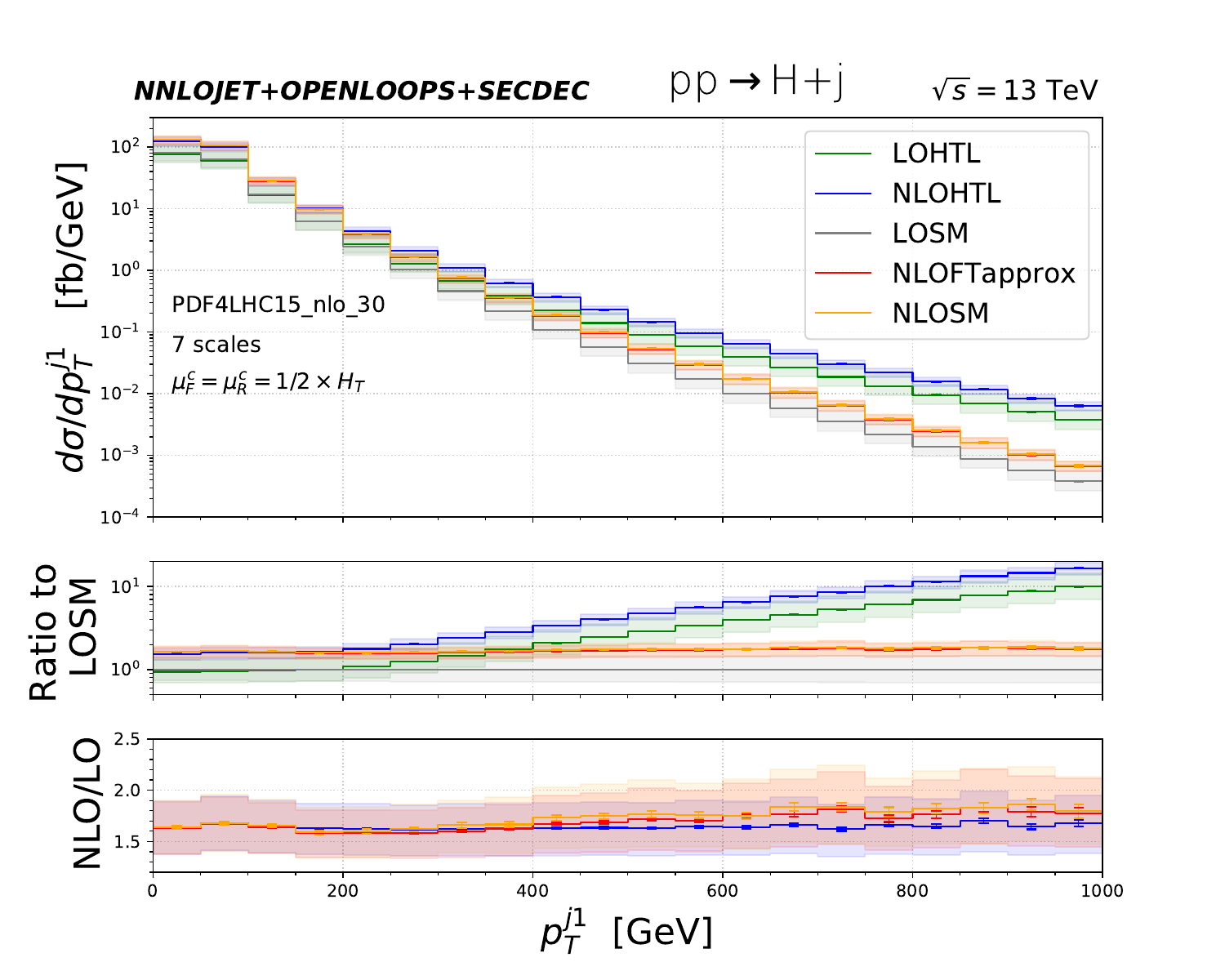}
\caption{Transverse momentum distribution of the Higgs (left) and the hardest jet (right) in \hj production. We show LO predictions in the full SM QCD (magenta) and the HTL (green) as well as NLO predictions in the \HTL (blue), the \FTapprox (red) and the full SM QCD (orange). The upper panel shows absolute predictions. The first ratio plot shows corrections with respect to LOSM, while the second ratio plot shows NLO corrections normalised to the respective LO prediction. Shaded bands correspond to scale variations. Error bars indicate integration uncertainties. (from Ref.~\cite{Chen:2021azt})}
\label{fig:hj_pth_ptj}
\end{figure}

\begin{figure}[t]
\centering
	\includegraphics[width=0.49\textwidth]{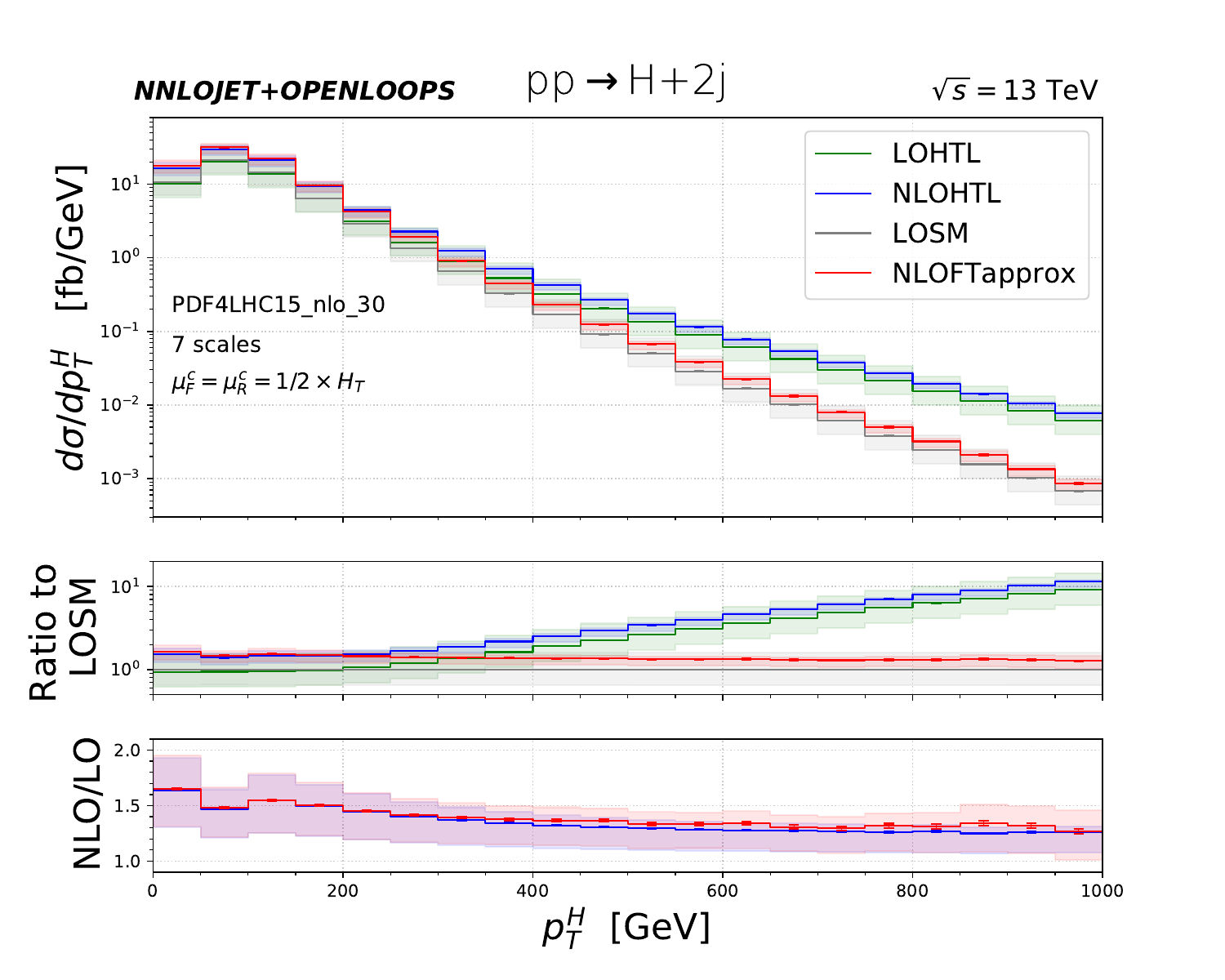}
\includegraphics[width=0.49\textwidth]{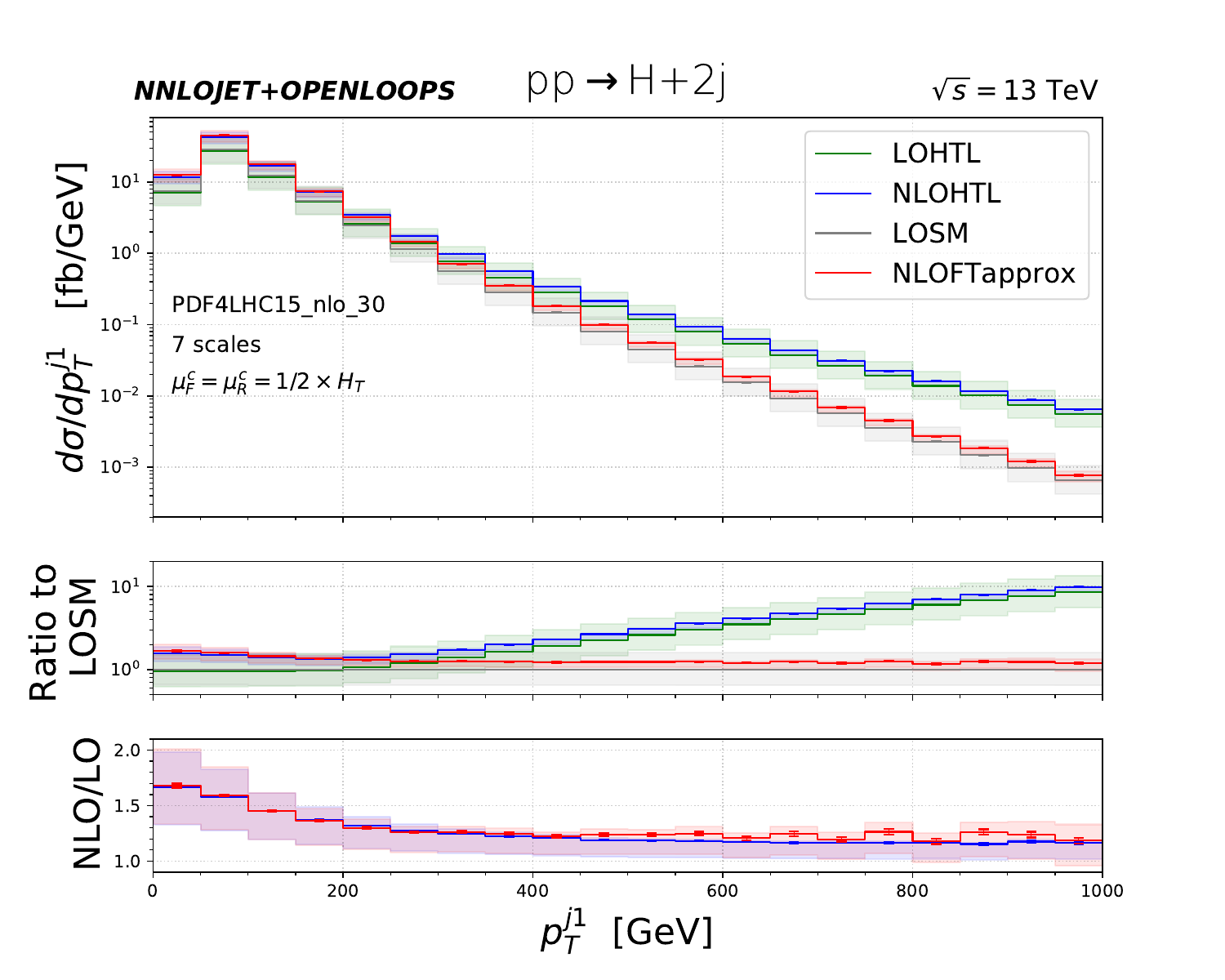}
\caption{Transverse momentum distribution of the Higgs (left) and the hardest jet (right) in \hjj production. Colour coding and labelling  as in Fig.~\ref{fig:hj_pth_ptj}. (from Ref.~\cite{Chen:2021azt})}
\label{fig:hjj_pth_ptj}
\end{figure}

We study top-quark effects for these processes through three approaches: 
the heavy top limit~(HTL),
the so-called full theory approximation~(\FTapprox) through a reweighting procedure based on the Born-level top-mass effects~\cite{Maltoni:2014eza},
and the exact top-mass dependence in the full SM QCD (only for \hj production).
We show the transverse momentum distribution for the Higgs boson ($p_T^H$) and the hardest jet ($p_T^{j1}$) for \hj production in Fig.~\ref{fig:hj_pth_ptj} and for \hjj production in Fig.~\ref{fig:hjj_pth_ptj}.
The fiducial total cross section for \hj and \hjj production are presented in Table~\ref{tab:xs}, including the results for boosted Higgs boson with $p_T^H > 300$~GeV.
A detailed study of fiducial total cross sections with different $p_T^H$  cuts ranging from 50~GeV to 800~GeV can be found in Ref.~\cite{Chen:2021azt}.

\renewcommand{\arraystretch}{1.5}
\begin{table}[t]
\centering
  \begin{tabular}{  ll|ccc | ccc }
    \multirow{2}{*}{\large$\sigma$[pb]} && \multicolumn{3}{c|}{Inclusive $p_{T,H}$} & \multicolumn{3}{c}{$p_{T,H}>300$ GeV}\\ 
    
              &   & LO & NLO & K & LO & NLO & K\\ \hline\hline
              
    \multirow{3}{*}{\hj} &HTL & $8.22^{+3.17}_{-2.15}$ & $13.57^{+2.11}_{-2.09}$  & $1.65$  & $0.086^{+0.038}_{-0.024}$ & $0.160^{+0.033}_{-0.030}$ & $1.86$  \\ 
     &\FTapprox & $8.56^{+3.30}_{-2.24}$ & $14.06(1)^{+2.17}_{-2.16}$ & $1.64$ & $0.046^{+0.020}_{-0.013}$ & $0.088^{+0.019}_{-0.017}$ & $1.91$ \\
     & SM QCD & $8.56^{+3.30}_{-2.24}$ &  $14.15(7)^{+2.29}_{-2.21}$ & $1.65$ & $0.046^{+0.020}_{-0.013}$ & $0.089(3)^{+0.020}_{-0.017}$ & $1.93$ \\    \hline
        \multirow{3}{*}{\hjj} &HTL & $2.87^{+1.67}_{-0.99}$ & $4.33^{+0.59}_{-0.80}$ & $1.51$ & $0.120^{+0.071}_{-0.042}$  & $0.160^{+0.012}_{-0.025}$ & $1.33$ \\
     &\FTapprox & $2.92^{+1.70}_{-1.01}$ & $4.45(1)^{+0.63}_{-0.83}$ & $1.52$ & $0.068^{+0.040}_{-0.024}$ & $0.092^{+0.008}_{-0.015}$ & $1.35$ \\
     & SM QCD & $2.92^{+1.70}_{-1.01}$ &  $-$ & $-$ & $0.068^{+0.040}_{-0.024}$ & $-$ & $-$   \end{tabular}
   \caption{\label{tab:totalXS}Integrated cross sections at LO and NLO QCD in the HTL and \FTapprox approximations and with full top-quark mass dependence (SM QCD) for \hj and \hjj production together with corresponding K-factors. Uncertainties correspond to the envelope of 7-point scale variations. For \hj production we require $p_{\rm{T},j}>30\,\GeV$, while for \hjj production we require $p_{\rm{T},j_1}>40\,\GeV,\, p_{\rm{T},j_2}>30\,\GeV$. On the left no further phase-space restrictions are considered, while on the right we require $p_{T,H}>300\,\GeV$. Numerical integration errors larger than permil level are indicated in brackets. (from Ref.~\cite{Chen:2021azt})
   }
   \label{tab:xs}
\end{table}

\section{NLO EW corrections for \hj and \hh production}
\begin{figure}[b]
  \centering
  \includegraphics[width=0.85\textwidth]{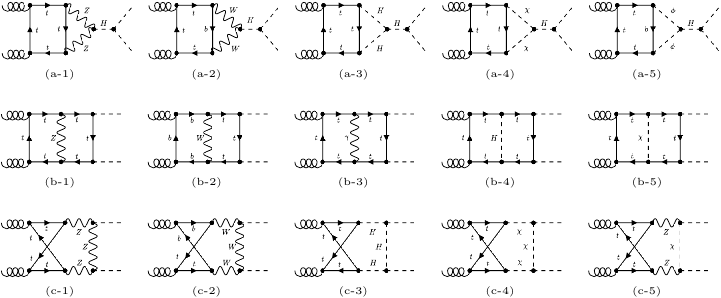}
  \caption{Two-loop Feynman diagrams contributing to $gg\to HH$.
  Dashed, solid, wavy and curly lines correspond to scalar particles,
  fermions, electroweak gauge bosons and gluons, respectively.}
  \label{fig::diags}
\end{figure}

We perform analytic two-loop calculations for NLO EW corrections for \hj and \hh production by expansions in both low-energy and high-energy regions. In particular, as shown in Fig.~\ref{fig:hj_pth_ptj}, the major contribution for \hj is from the low-energy region, which can be captured by a large-$m_t$ expansion. 
A dedicated review on analytic expansion for EW corrections can be found in Ref.~\cite{Zhang:2023our}. 
Note that the factorisable contributions for $gg\to HH$ are computed in a fully analytic form in Ref.~\cite{Zhang:2024rix}.
Sample two-loop diagrams
contributing to $gg\to HH$ are shown in Fig.~\ref{fig::diags}.
\begin{figure}[t]
  \centering
  \begin{tabular}{cc}
    \includegraphics[width=.4\textwidth]{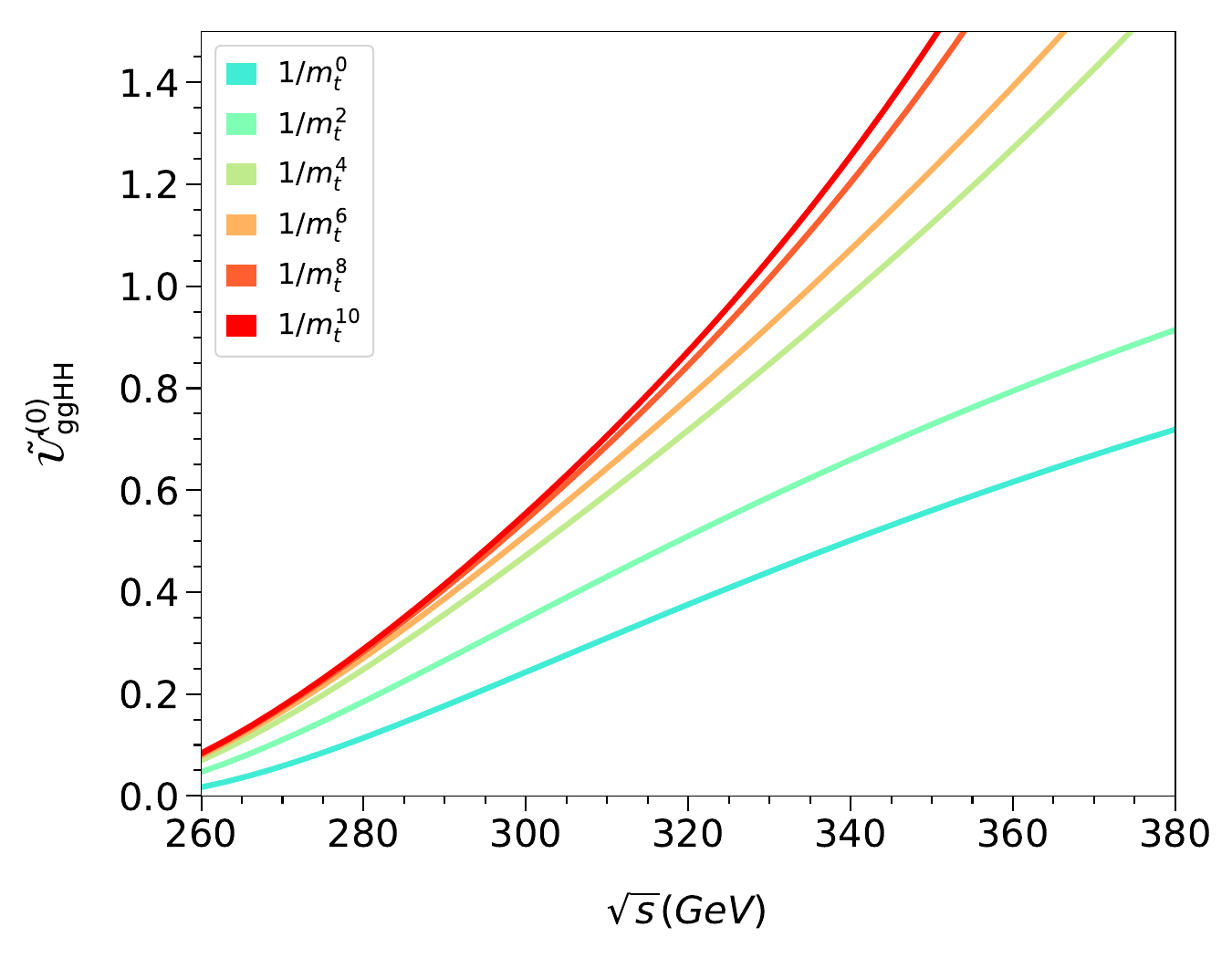} &
    \includegraphics[width=.4\textwidth]{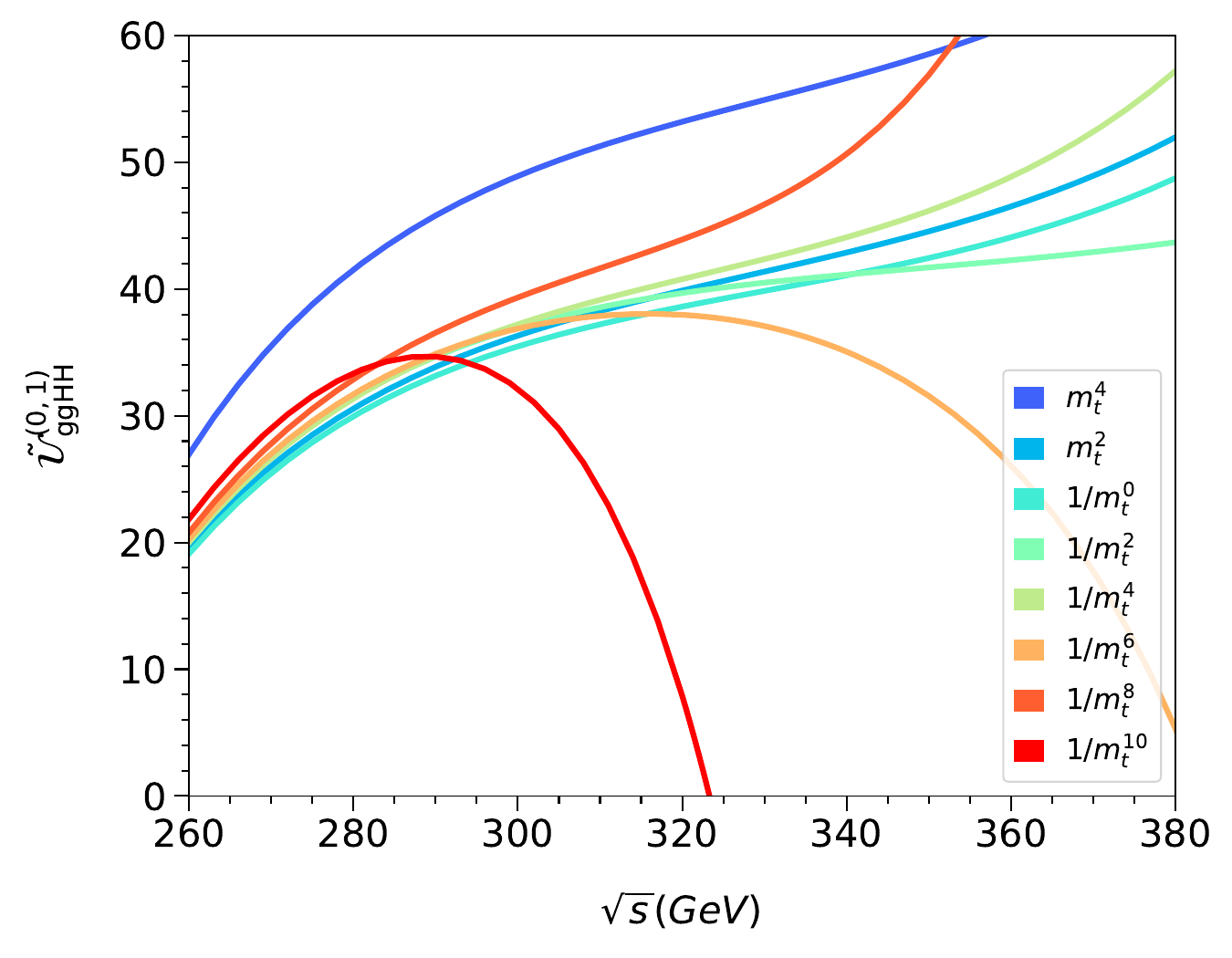}
  \end{tabular}
  \caption{\label{fig::gghh}
      LO $\tilde{{\cal U}}_{\rm ggHH}^{(0)}$ (left) and NLO EW $\tilde{{\cal U}}_{\rm ggHH}^{(0,1)}$ (right) matrix elements plotted as a function of
      $\sqrt{s}$ for $gg\to HH$. Results are shown up to order $1/m_t^{10}$ (from Ref.~\cite{Davies:2023npk}).
       }
\end{figure}
\begin{figure}[t!]
  \centering
  \begin{tabular}{cc}
    \includegraphics[width=.4\textwidth]{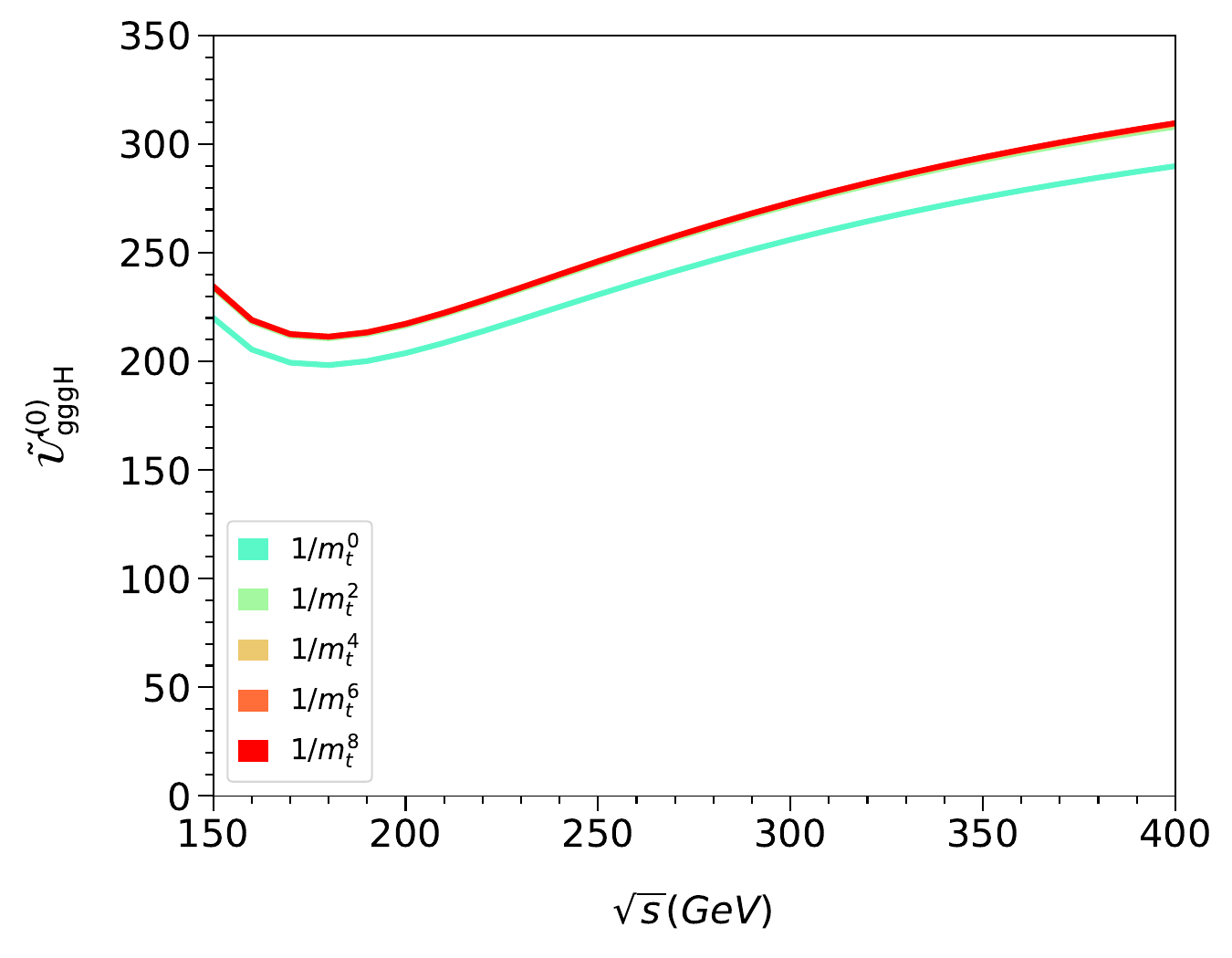} &
    \includegraphics[width=.4\textwidth]{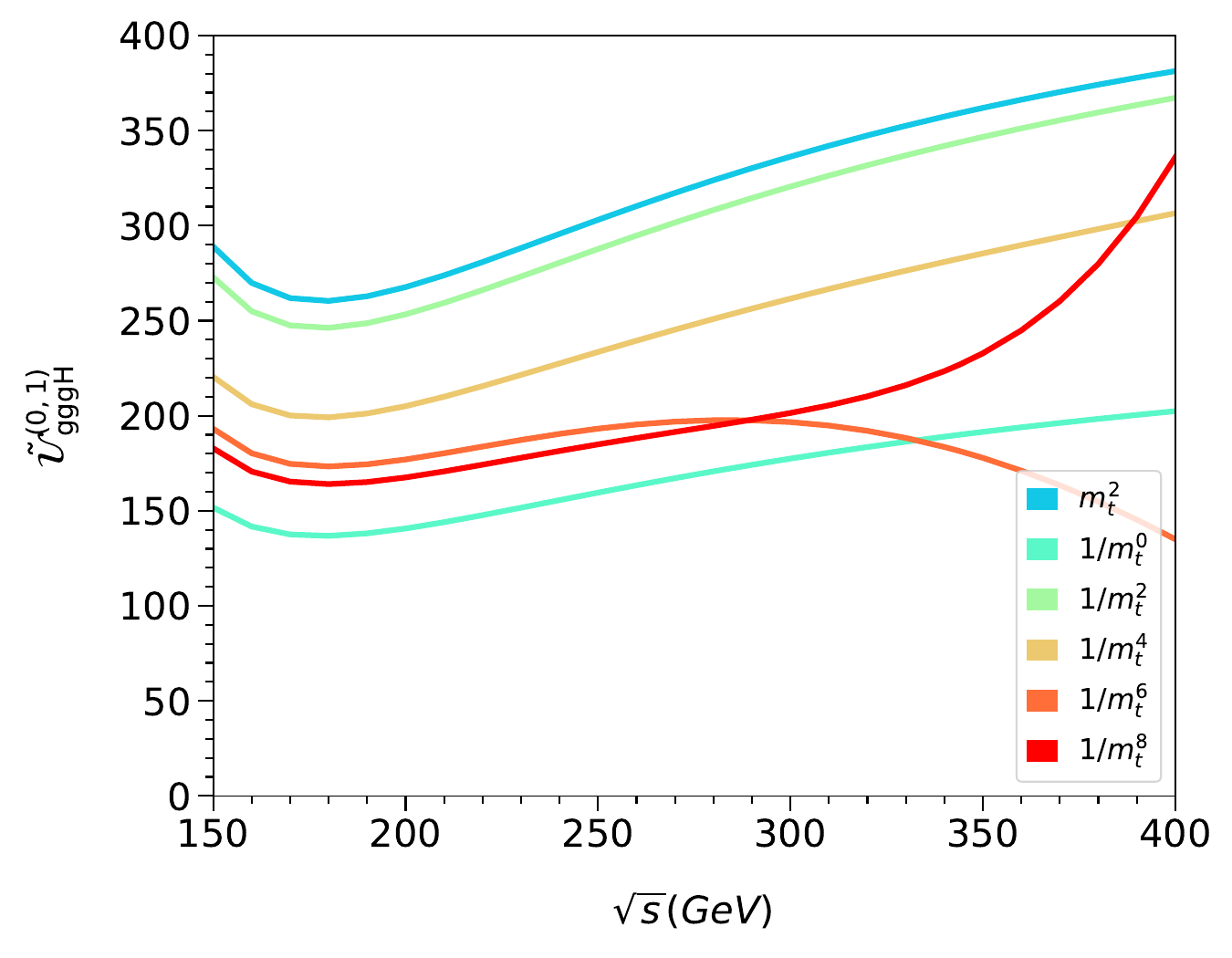}
  \end{tabular}
  \caption{\label{fig::gggH}
  LO $\tilde{{\cal U}}_{\rm gggH}^{(0)}$ (left) and NLO EW $\tilde{{\cal U}}_{\rm gggH}^{(0,1)}$ (right) matrix elements plotted as a function of
      $\sqrt{s}$ for $gg\to gH$. Results are shown up to order $1/m_t^{8}$. (from Ref.~\cite{Davies:2023npk})
    }
\end{figure}

\paragraph{Large-$m_t$ expansion:}

We perform analytic calculations for the full top-quark-induced EW corrections for gluon-fusion \hj  ($gg\to gH$) and \hh ($gg\to HH$) production 
through the large-$m_t$ expansion.
We assume the expansion hierarchy 
\begin{eqnarray}
  m_t^2 \gg \xi_W m_W^2, \xi_Z m_Z^2 \gg s,t,m_W^2,m_Z^2, m_H^2 \,,
  \label{eq::hierarchy-xi}
\end{eqnarray}
where $\xi_Z$, $\xi_W$ are the gauge-fixing parameters.
The large-$m_t$ expansion for bare two-loop amplitudes is performed to the order $1/m_t^8$ with the help of {\tt exp}~\cite{Harlander:1998cmq}.
We perform parameter renormalisation for independent parameters $  \{ e, m_W, m_Z, m_t, m_H \}$ with $e = \sqrt{4\pi\alpha}$
and external Higgs fields renormalisation, both in the on-shell scheme ($G_\mu$ scheme).
Note that $\xi_W$ and $\xi_Z$ drop out for both $gg\to HH$ and $gg\to gH$ renormalised amplitudes.

For the numerical evaluation, we use the inputs
$
  m_t = 172~\mbox{GeV}\,,  m_H = 125~\mbox{GeV}\,,
  m_W = 80~\mbox{GeV}\,, m_Z = 91~\mbox{GeV}\,,
$
and introduce the ratio parameter
$
  \rho_{p_T} = \frac{p_T^H}{\sqrt{s}}
$.
In the following we choose $\rho_{p_T}=0.1$ and
consider the squared matrix element 
$  {\cal U}_{\rm ggHH} =
{\cal U}_{\rm ggHH}^{(0)}
+\frac{\alpha}{\pi} {\cal U}_{\rm ggHH}^{(0,1)}
$
 for $gg\to HH$ and
$  {\cal U}_{\rm gggH} =
{\cal U}_{\rm gggH}^{(0)}
+\frac{\alpha}{\pi} {\cal U}_{\rm gggH}^{(0,1)}
$
 for $gg\to gH$.
For definitions of matrix elements and their prefactors, please refer to Ref.~\cite{Davies:2023npk}.
The LO and NLO EW numerical results are shown in Fig.~\ref{fig::gghh} for $gg\to HH$
and Fig.~\ref{fig::gggH} for $gg\to gH$.
For both processes, our large-$m_t$ expansions yield reasonable predictions for the $\sqrt{s} \lsim 290$~GeV region.
We observe that the EW corrections for $gg\to HH$ are sizeable in the low-energy region, lifting the di-Higgs destructive interference effects which is present at the LO.
The EW corrections for $gg\to gH$ are relatively small.
%

\begin{figure}[t]
  \centering
  \begin{tabular}{cc}
    \includegraphics[width=0.4\textwidth]{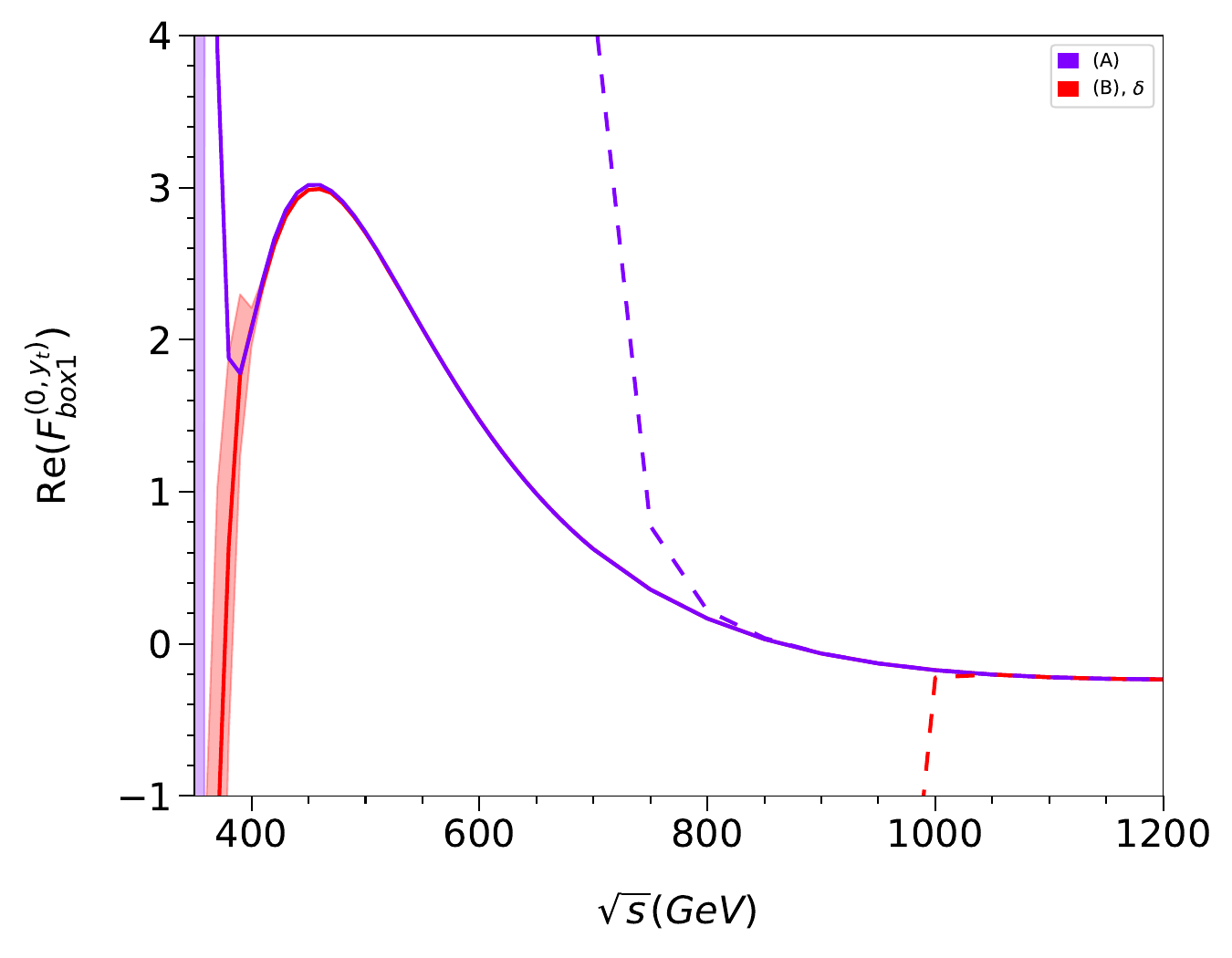}
    &
    \includegraphics[width=0.4\textwidth]{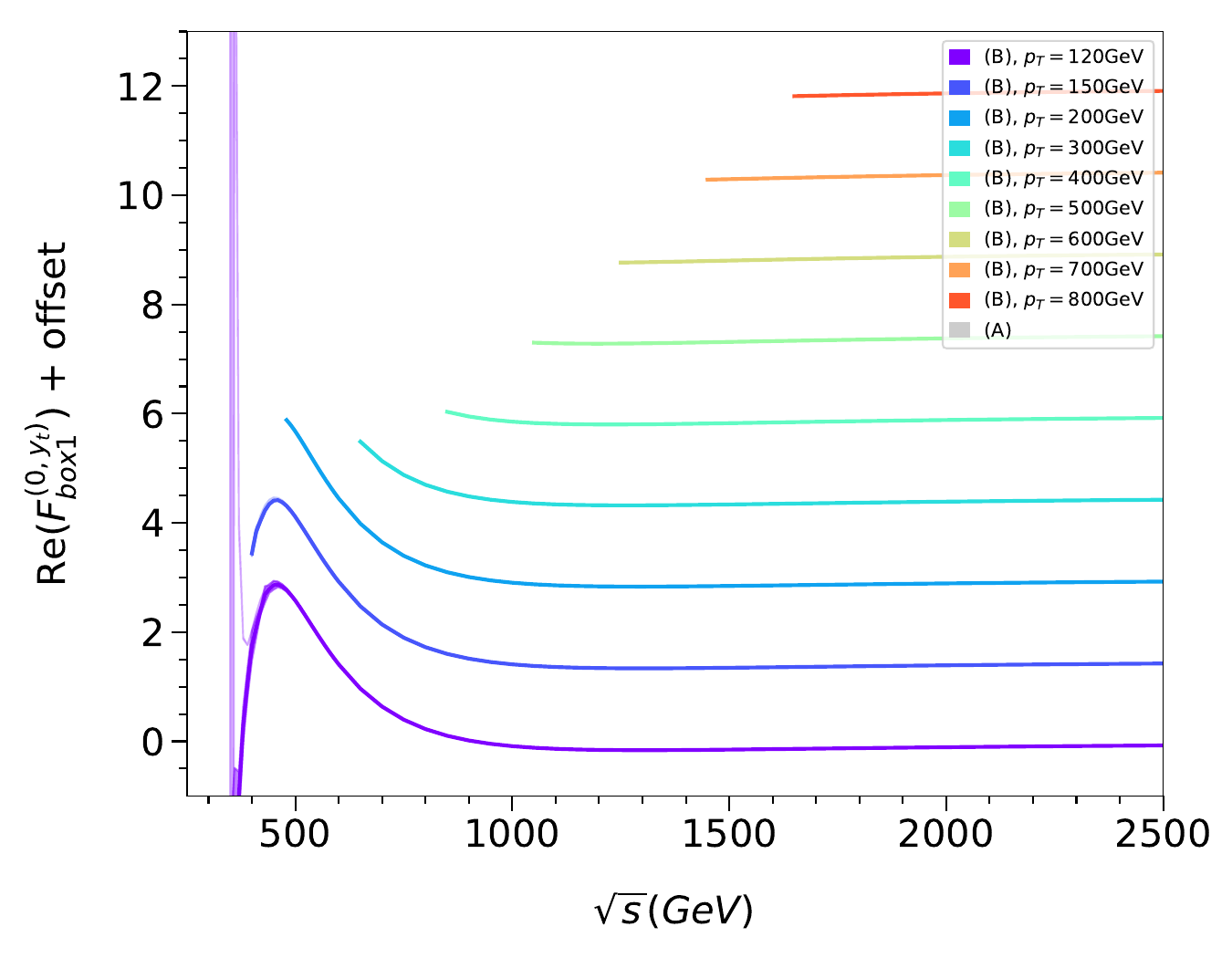}
  \end{tabular}
  \caption{\label{fig::F12}
  Real part of $F_{\rm box1}$ for fixed scattering angle $\theta=\pi/2$ (left) and different fixed $p_T$ values (right). 
  For the fixed $p_T$ plot, an offset is applied 
    such that the curves with different $p_T$ are separated.
(from Ref.~\cite{Davies:2022ram})
    }
\end{figure}
\paragraph{High-energy expansion:}
We perform the analytic high-energy expansion for the two-loop leading Yukawa corrections for \hh production.
This contribution is generated through diagrams of the type (b-4) shown in Fig.~\ref{fig::diags}.
We consider two approaches with the following hierarchies:
\begin{eqnarray}
({\rm A}) \,\; s,t \gg m_t^2 \gg (m_H^{\rm int})^2, (m_H^{\rm ext})^2 \quad \mbox{and} \quad 
({\rm B}) \,\; s,t \gg m_t^2 \approx (m_H^{\rm int})^2 \gg (m_H^{\rm ext})^2\,,
\end{eqnarray}
where $m_H^{\rm int}$ and $m_H^{\rm ext}$ are internal and external Higgs masses respectively.
We employ {\tt exp} for amplitude-level expansions, 
differential equations and power-log ansatz methods for master integrals, 
and compute the boundary master integrals in the high-energy limit using \texttt{AsyInt}~\cite{Zhang:2024fcu}.
Through this procedure, we obtain analytic results for the amplitudes expanded up to order $m_t^{120}$.
For the numerical evaluation, we employ the Pad\'{e} approximation to enlarge the radius of convergence of our results.
We show results for the real part of box-type form factor of $F_{1}$ for
fixed transverse momentum $p_T$ and fixed scattering angle in Fig.~\ref{fig::F12}.
For the fixed scattering angle plot,
the solid curves represent Pad\'e results with uncertainty bands
and the dashed curves show naive expansions.
We observe the central value of Pad\'e results agree in both approaches down to $\sqrt{s}\approx 400$~GeV.
%

\section{Conclusion}
In these proceedings, we have summarised our recent contributions to NLO QCD and EW corrections for loop-induced Higgs bosons production at the LHC. These include the QCD and EW corrections for \hj production, QCD corrections for \hjj production, and EW corrections for \hh production.
For QCD corrections, we have conducted dedicated studies on the top-quark mass effects for \hj and \hjj production.
These effects are crucial for making precise predictions of loop-induced Higgs boson production at the NLO accuracy. 
For EW corrections, we have performed analytic expansions for \hj and \hh production in different kinematic regions.
These results provide a good approximation across a vast range of phase space regions, showing a promising prospects for analytic higher-order EW calculations.

\section*{Acknowledgments}
H.Z. would like to thank Joshua Davies, Go Mishima, Kay Sch\"onwald, Matthias Steinhauser, Xuan Chen, Alexander Huss, Stephen Jones, Matthias Kerner, Jean-Nicolas Lang, and Jonas Lindert for collaborations on projects reported in this contribution.
H.Z. would like to thank the organisers of ICHEP2024 for the possibility to present our results at the conference. 
H.Z. is supported by the Deutsche Forschungsgemeinschaft (DFG, German Research Foundation) under grant 396021762 --- TRR 257 “Particle Physics Phenomenology after the Higgs Discovery”.

\bibliographystyle{JHEPmod}
\small
\setlength{\bibsep}{3pt}
\bibliography{reference}

\end{document}